\documentclass[sigconf]{acmart}

\usepackage{url}
\usepackage{booktabs} 
\usepackage{graphicx}

\usepackage{algorithm}
\usepackage[noend]{algpseudocode}
\usepackage{xcolor, soul}

\definecolor{lgray}{rgb}{.90,.90,0.90}

\usepackage{colortbl}

\definecolor{lblue}{rgb}{0.7, 0.7, 1}

\begin{document}

\newcommand{\kmm}[1]{\emph{[#1 -kmm]}}


\addtolength{\abovedisplayskip}{-1.15em}
\addtolength{\belowdisplayskip}{-1.65em}
\addtolength{\textfloatsep}{-1em}
\addtolength{\intextsep}{-0.5em}
\addtolength{\parskip}{-0.38pt}
\addtolength{\baselineskip}{-0.38pt}

\copyrightyear{} 
\acmPrice{ }
\acmYear{ }
\setcopyright{none}
\acmConference[RecSys 2017 Poster Proceedings, August 27-31, Como, Italy.]{}{}

\title{Explainable Entity-based Recommendations\\ with Knowledge Graphs}

\author{\textbf{Rose Catherine \hspace{.5cm} Kathryn Mazaitis  \hspace{.5cm} 	Maxine Eskenazi   \hspace{.5cm}  William Cohen}}
\affiliation{%
  \institution{School of Computer Science \\ Carnegie Mellon University \\
  \texttt{\{rosecatherinek,krivard,max,wcohen\}@cs.cmu.edu}}
}

%

\begin{abstract}
Explainable recommendation is an important task. 
Many methods have been proposed which generate explanations from the content and reviews written for items. 
When review text is unavailable, generating explanations is still a hard problem. 
In this paper, we illustrate how explanations can be generated in such a scenario by leveraging external knowledge in the form of knowledge graphs. Our method jointly ranks items and knowledge graph entities using a Personalized PageRank procedure to produce recommendations together with their explanations. 
\end{abstract}


\maketitle

\section{Introduction}
Improving the accuracy of predictions in recommender systems is an important research topic. An equally important task is explaining the predictions to the user. 
Explanations may serve many different purposes \cite{eval}. They can show how the system works (transparency) or help users make an informed choice (effectiveness). They may be evaluated on whether they convince the user to make a purchase (persuasiveness) or whether they help the user make decision more rapidly (efficiency). In general, providing an explanation has been shown to build user's trust in the recommender system \cite{trust}. 

The focus of this paper is a system to generate explanations for Knowledge Graph (KG) -based recommendation. Users and items are typically associated with factual data, referred to as \textit{content}. For users, the content may include demographics and other profile data. For items such as movies, it might include the actors, directors, genre, and the like. The KG encodes the interconnections between such facts, and leveraging these links has been shown to improve recommender performance \cite{rose_recsys16, heterec, nell}.

Although a number of explanation schemes have been proposed in the past (Section \ref{sec_rel}), there has been no work which produces explanations for KG-based recommenders. In this paper, we present a method to jointly rank items and entities in the KG such that the entities can serve as an explanation for the recommendation. 

Our technique can be run without training, thereby allowing faster deployment in new domains. Once enough data has been collected, it can then be trained to yield better performance. 
The proposed method can also be used in a dialog setting, where a user interacts with the system to refine its suggestions. 

\section{Related Work}
\label{sec_rel}

Generating explanations for recommendations has been an active area of research for more than a decade. 
\cite{exp_cf} was an early work  that assessed different ways of explaining recommendations in a collaborative filtering (CF) -based recommender system. 

In content-based recommenders, the explanations revolve around the content or profile of the user and the item. 
The system of \cite{kw_book} simply  displayed keyword matches between the user's profile and the books being recommended. 
Similarly, \cite{tags_mov} proposed a method called `Tagsplanations', which showed the degree to which a tag is relevant to the item, and the sentiment of the user towards the tag. 

With the advent of social networks, explanations that leverage social connections have also gained attention. For example, \cite{exp_social} produced explanations that showed whether a good friend of the user has liked something, where friendship strength was computed from their interactions on Facebook. 

More recent research has focused on providing explanations that are extracted from user written reviews for the items. 
\cite{review1} extracted phrases and sentiments expressed in the reviews and used them to generate  explanations. 
\cite{HFHT} uses topics learned from the reviews as aspects of the item, and uses the topic distribution in the reviews to find useful or representative reviews. 

Knowledge Graphs have been shown to improve the performance of recommender systems in the past. 
\cite{heterec} proposed a meta-path based method that learned paths consisting of node types in a graph. 
Similarly, \cite{sprank}  used paths to find the top-N recommendations in a learning-to-rank framework.  
A few methods such as \cite{nell, lod} rank items using Personalized PageRank. In these methods, the entities present in the text of an item 
are first mapped to entities in a knowledge graph.
\cite{rose_recsys16} proposed  probabilistic logic programming models for  recommendation on knowledge graphs. 
None of the above KB-based recommenders attempted to generate explanations.

\section{Explanation Method}
\label{sec_prop}

In this section, we propose our method, which builds on the work of  \cite{rose_recsys16} by using ProPPR \cite{proppr1} for learning to recommend. ProPPR (\textbf{Pro}gramming with \textbf{P}ersonalized \textbf{P}age \textbf{R}ank) is a first order logic system. It takes as input a set of rules and a database of facts, and uses these to generate an approximate local grounding of each query in a small graph. Candidate answers to the query are the nodes in the graph that satisfy the rules. The candidates are then ranked by running a Personalized PageRank algorithm on the  graph. 

Our technique proceeds in two main steps. First, it uses ProPPR to jointly rank items and entities for a user. Second, it consolidates the results into recommendations and explanations. 

To use ProPPR to rank items and entities, we first define a notion of similarity between nodes in the graph, using the same similarity rules as \cite{rose_recsys16} (Figure \ref{rule_sim}). This simple rule states that two entities \texttt{X} and \texttt{E} are similar if they are the same (Rule 1), or if there is a link in the graph connecting \texttt{X} to another entity \texttt{Z}, which is similar to \texttt{E} (Rule 2). Note that this definition of similarity is recursive.

\begin{figure}[htb]
\centering
\begin{align}
\mathtt{sim(X,X)} \leftarrow & \mathtt{true.} \\
 \mathtt{sim(X,E)} \leftarrow & \mathtt{link(X,Z), sim(Z,E).} 
\end{align}
\caption{Similarity in a graph}
\label{rule_sim}
\end{figure}

Next, the model has two sets of rules for ranking: one set for joint ranking of movies that the user would like, together with the most likely reason (Figure \ref{rule_likes}), and a similar set for movies that the user would not like (Figure \ref{rule_dislikes}).
In Figure \ref{rule_likes}, Rule 3 states that a user \texttt{U} will like an entity \texttt{E} and a movie \texttt{M} if the user likes the entity, and the entity is related (\texttt{sim}) to the movie. The clause \texttt{isMovie} ensures that the variable \texttt{M} is bound to a movie, since \texttt{sim} admits all types of entities.
Rule 3 invokes the predicate \texttt{likes(U,E)}, which holds for an entity \texttt{E} if the user has explicitly stated that they like it (Rule 4), or if they have provided positive feedback (e.g. clicked, thumbs up, high star rating) for a movie \texttt{M} containing (via \texttt{link(M,E)}) the entity (Rule 5). 
%
The method for finding movies and entities that the user will dislike is  similar to the above, as given in Figure \ref{rule_dislikes}.

\begin{figure}[htb]
\centering
\begin{align}
\mathtt{willLike(U,E,M)} \leftarrow & \mathtt{likes(U,E), sim(E,M), isMovie(M).} \\
 \mathtt{likes(U,E)} \leftarrow & \mathtt{likesEntity(U,E).} \\
\mathtt{likes(U,E)} \leftarrow & \mathtt{likesMovie(U,M),link(M,E).}
\end{align}
\caption{Predicting likes}
\label{rule_likes}
\end{figure}

 \begin{figure}[htb]
 \centering
 \begin{align}
 \mathtt{willDislike(U,E,M)} \leftarrow & \mathtt{dislikes(U,E), sim(E,M), isMovie(M).} \nonumber\\
  \mathtt{dislikes(U,E)} \leftarrow & \mathtt{dislikesEntity(U,E).}  \nonumber\\
 \mathtt{dislikes(U,E)} \leftarrow & \mathtt{dislikesMovie(U,M),link(M,E).}  \nonumber
 \end{align}
 \caption{Predicting Dislikes}
 \label{rule_dislikes}
 \end{figure}

\begin{figure}[htb]
\centering
\includegraphics[bb = 20 140 794 605, clip, width=0.49\textwidth]{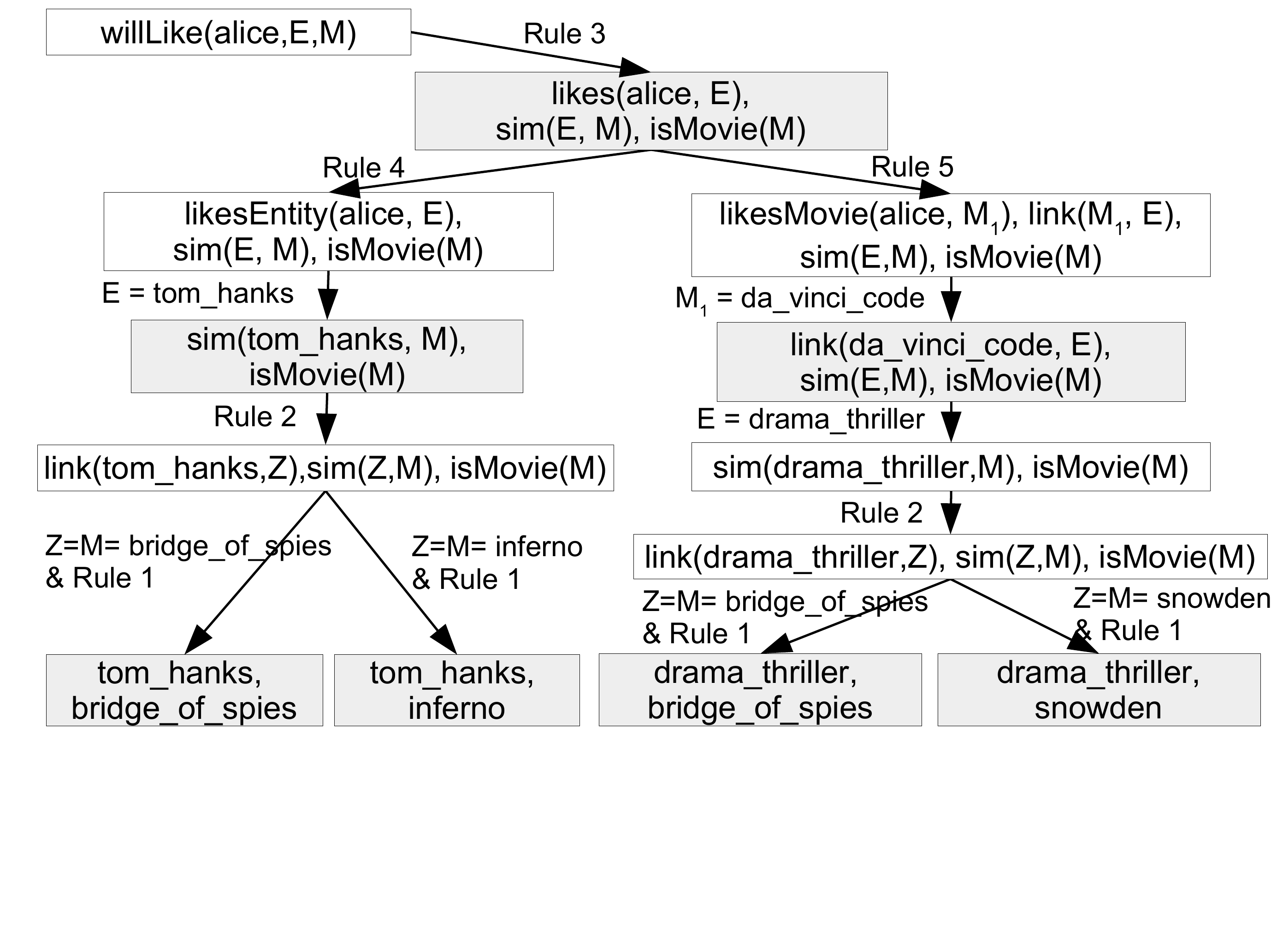}
\caption{Sample grounding for predicting likes}
\label{pprprove}
\end{figure}

To jointly rank the items and entities, we use ProPPR to query the \texttt{willLike(U,E,M)} predicate with the user specified and the other two variables free. Then, the ProPPR engine will ground the query into a proof graph by replacing each variable recursively with literals that satisfy the rules from the KG \cite{rose_recsys16, proppr1}. A sample grounding when queried for a user \texttt{alice} who likes \texttt{tom\_hanks} and the movie \texttt{da\_vinci\_code}  is shown in Figure \ref{pprprove}. 

After constructing the proof graph, ProPPR runs a Personalized PageRank algorithm with \texttt{willLike(alice, E, M)} as the start node. In this simple example, 
we will let the scores for \texttt{(tom\_hanks, bridge\_of\_spies)}, \texttt{(tom\_hanks, inferno)}, \texttt{(drama\_thriller, bridge\_of\_spies)}, and \texttt{(drama\_thriller, snowden)}, be 0.4, 0.4, 0.3 and 0.3 respectively. 

Now, let us suppose that \texttt{alice} has also specified that she dislikes \texttt{crime} movies. If we follow the grounding procedure for dislikes 
and rank the answers, we may obtain \texttt{(crime, inferno)} with score 0.2. Our system then proceeds to consolidate the recommendations and the explanations by grouping by movie names, adding together their `like' scores and deducting their `dislike' scores. For each movie, the entities can be ranked according to their joint score. The end result is a list of reasons which can be shown to the user:  
\begin{sloppypar}
\begin{enumerate}
\item \texttt{bridge\_of\_spies}, score = 0.4 + 0.3 = 0.7, reasons = \\\{ tom\_hanks, drama\_thriller \}
\item \texttt{snowden}, score = 0.3, reasons = \{ drama\_thriller \}
\item \texttt{inferno}, score = 0.4 - 0.2 = 0.2, reasons = \{ tom\_hanks, (-ve) crime \}
\end{enumerate}
\end{sloppypar}


\section{Real World Deployment}

The proposed method  is presently used as the backend of a personal agent running on mobile devices for recommending  movies \cite{inmind} undergoing Beta testing. The knowledge graph for recommendations is constructed from the weekly dump files released by \url{imdb.com}. The personal agent uses a dialog model of interaction with the user. In this setting, users are actively involved in refining the recommendations depending on what their mood might be. For example, for a fun night out with friends, a user may want to watch an action movie, whereas when spending time with her significant other, the same user may be in the mood for a romantic comedy. 

\section{Conclusions}
\label{sec_concl}
Knowledge graphs have been shown to improve recommender system accuracy in the past. However, generating explanations to help users make an informed choice in KG-based systems has not been attempted before. In this paper, we proposed a method to produce a ranked list of entities as explanations by jointly ranking them with the corresponding movies.

\bibliographystyle{ACM-Reference-Format}
\bibliography{mybib} 

\end{document}